# MATLAB Scripts for RF Commissioning at the LANSCE LINAC*

Sungil Kwon[†], A. Archuleta, L. Castellano, C. Marchwinski, M. Prokop, P. Van Rooy, P. Torrez
Los Alamos National Laboratory, Los Alamos, NM, USA

*Abstract*

The linear accelerator (LINAC) at the Los Alamos Neutron Science Center (LANSCE) consists of Pre-buncher, Main-Buncher, low-energy beam transport (LEBT), four 201.25-MHz Drift Tube Linacs (DTLs) and forty-four 805-MHz Coupled Cavity Linacs (CCLs). As a part of the upcoming LANSCE Modernization project, low-level RF (LLRF) systems of four 201-MHz DTLs and twenty-six 805-MHz SCLs have been digitized. Hence the network-based control of the cavity field and RF commissioning are possible. Each LLRF and high-power RF (HPRF) systems have many process variables (PVs) located on different computer control screens provided by the Extensible Display Manager (EDM). Several MATLAB m-scripts have been developed to efficiently process the necessary PVs while auto-start, amplitude/phase calibration, gain tuning of the cavity field feedback controllers, gain and phase tuning of the beam feedforward controllers, and high power RF trip recovery, processes are configured and validated. This paper addresses the sequence of RF commissioning of the LANSCE LINAC from the time of RF-turn-on to beam feedforward control and its relevant EDMs and MATLAB m-scripts.

## 1. INTRODUCTION

The LANSCE LINAC is composed of one 201.25 MHz Drift-Tube_Linac (DTL) sector and seven 805MHz Coupled-Cavity Linac(CCL) sectors. A 805 MHz sector is composed of 6~7 CCLs with one high voltage system, Capacity Bank. In contrast, the 201.25 MHz sector is composed of 4 DTLs and each DTL has its own high voltage system [1]. The LLRF systems of four DTLs and twenty-six CCLs were digitized in the last several years.

Since these digital LLRF (DLLRF) systems work in the baseband frequency domain, the structure of the LLRF control system are the same and the digital signal processing (DSP) – downconversion to the baseband signals, filtering, control signal generation, upconversion to the IF signals – that are implemented on the Intel FPGA are very similar. Figure 1 shows the overall LLRF control system block diagram. In figure 1, all submodules such as the RF upconverter, intermediate power amplifier and high power amplifier, on the RF drive path are represented as the lumped transfer function $H_{fwd}(s)$ and all submodules such as RF pickup, RF downconverter, IF filter, on the RF feedback path is represented as the lumped transfer function $H_{ret}(s)$ [2].

Because the digitization was accompanied with the LANSCE Control System (LCS) network support, the network based control of the LLRF control system became possible. Control parameters of LLRF system are assigned with parameter variables (PVs) of the EPICS Database and the operation of the (LL)RF system can be performed by adjusting those PVs properly. In addition, since the measured signals of the LLRF control system are transmitted to the LCS network, it is possible to implement data based digital signal processing on the host system connected to the LCS network.

MATLAB is used widely at LANSCE for signal/image analysis and synthesis of LLRF digital signal processing (DSP). MATLAB is also used for controlling PVs of (LL)RF system. For this, a compact MATLAB version of EPICS channel access (CA) was developed, where MATLAB m-script uses Operating System (OS) function "*system*" to call EPICS CA functions, *caget, caput*, etc..

The procedure of the RF operation and Beam operation is comprised of (i) turn on RF and adjust amplitude and phase set points of the cavity field to the pre-assigned target values; (ii) calibrate the open loop gain and phase; (iii) close the PI feedback loop and calibrate the cavity amplitude and the cavity phase of the external measurements to the stored target values; (iv) tune the beam feedforward controller. Automations of (i), (iii), (iv) are accompanied with callings of MATLAB functions.

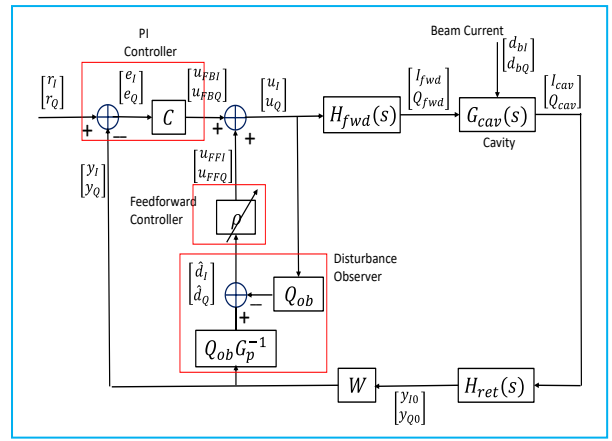

Figure 1. LLRF Control System Block Diagram

## 2. RF TURN ON FOR START UP OR VSWR RECOVERY

RF turn on and adjustments of amplitude and phase set points are performed either for the first start up of the LINAC RF operation or for the RF Trip Recovery. Target amplitude and phase set points are retrieved from an au-tosaved data file. The amplitude set point increment rate is pre-assigned. While the amplitude set point is increased, the reflected power and the klystron power are monitored to prevent VSWR(RF trip). The reflected power threshold is pre-assigned from the operational experience. When the amplitude set point is increased, the monitored reflected power may be over the threshold. In that case, the amplitude set point increment rate is adjusted by applying a fuzzy logic.

___________________
* Work supported by U.S. Dept. of Energy
† email address: skwon@lanl.gov.

This procedure is coded in a MATLAB function (**olpstartup.m**) and can be performed for one Module at a time or for part or all modules in a sector. Figure 2 shows the results of simultaneous startups of 2 LLRF systems in a 805MHz sector. The call to the MATLAB function for the startup is implemented on EPICS EDM screen (Figure 3) with Shell Command of EDM (figure 4).

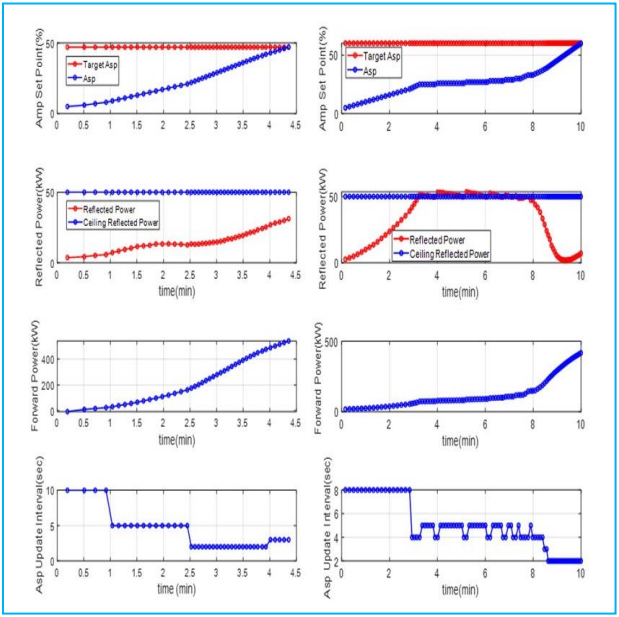

Figure 2. PV trajectories while RF turn on procedure for 3 Modules are performed simultaneously. (Left Col.: M6, right col.: M7).

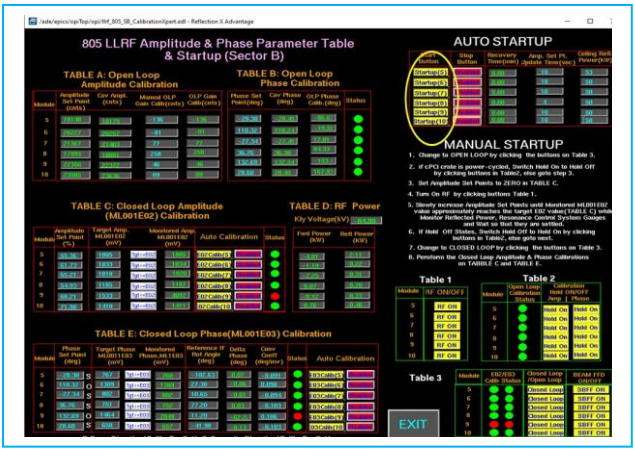

Figure 3. LLRF System Status Monitoring EDM Screen where Shell Command Button calls **olpstartup.m.**

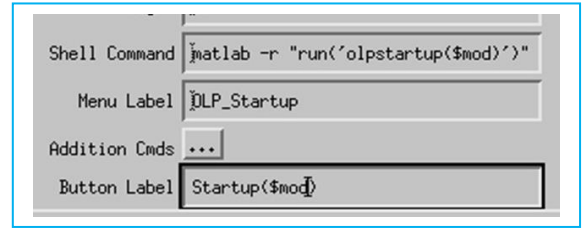

Figure 4. EDM Shell Command that calls the MATLAB function **olpstartup**($module)

## 3. DECOUPLING CONTROLLER

The accelerator RF cavities are modelled as two-input-two-output (TITO) systems. In the Laplace domain, the LLRF system can be represented with the nominal plant $H_n(s)$ and the multiplicative uncertainty $\Delta(s, \Delta\omega)$ [2,3] as

$$H_p(s) = H_n(s)(I + \Delta(s, \Delta\omega)V_\Delta(s)) \quad (1)$$

$$H_n(s) = \frac{h}{\tau_p s + 1}\begin{bmatrix} cos(\theta) & -sin(\theta) \\ sin(\theta) & cos(\theta) \end{bmatrix} \quad (2)$$

$$\|\Delta(j\omega, \Delta\omega)\|_\infty < 1 \quad (3)$$

In (1)-(3), $\Delta\omega$ is the detuning frequency, $\tau_p$ is the time constant of the cavity, $h$ is the steady state loop gain, $\theta$ is the overall phase rotation of the loop, $V_\Delta(s)$ is a shaping filter satisfying

$$\|\Delta(j\omega, \Delta\omega)\|_\infty \leq \|V_\Delta(j\omega)\|_\infty$$

For the nominal plant $H_n(s)$, a decoupling controller $W(s)$ given by

$$W(s) = \frac{1}{h}\begin{bmatrix} cos(\theta) & sin(\theta) \\ -sin(\theta) & cos(\theta) \end{bmatrix} \quad (4)$$

achieves the zero crosstalks between I/Q channels. However, applying (4) to the perturbed plant $H_p(s)$ yields another coupled, perturbed plant $G_p(s)$

$$G_p(s) = G_n(s)(I + \Delta(s, \Delta\omega)V_\Delta(s)), \quad (5)$$

in which the off-diagonal terms of $\Delta(s, \Delta\omega)$ are nonzero unless $\Delta\omega$ is zero. This problem of the model based decoupling controller design can be solved by a signal based approach of the decoupling controller design.

Let the amplitude and phase set point trajectories be $A_r(t)$ and $\theta_r(t)$ and let the measured amplitude and phase output trajectories be $A_{y0}(t)$ and $\theta_{y0}(t)$ which are also the inputs to the decoupling controller of the form

$$W(s) = g_w \begin{bmatrix} cos(\varphi) & -sin(\varphi) \\ sin(\varphi) & cos(\varphi) \end{bmatrix}. \quad (6)$$

When $W(s)$ is applied to $A_{y0}(t)$ and $\theta_{y0}(t)$, as shown in figure 1, the outputs $y_I(t)$ and $y_Q(t)$ of the decoupling controller become

$$\begin{bmatrix} y_I(t) \\ y_Q(t) \end{bmatrix} = W(s) \begin{bmatrix} y_{I0}(t) \\ y_{Q0}(t) \end{bmatrix}$$
$$= g_w A_{y0}(t) \begin{bmatrix} \cos(\varphi + \theta_{y0}(t)) \\ \sin(\varphi + \theta_{y0}(t)) \end{bmatrix}. \quad (7)$$

The purpose of the decoupling controller design is that the right-side equation of (7) satisfies

$$g_w A_{y0}(t) \begin{bmatrix} \cos(\varphi + \theta_{y0}(t)) \\ \sin(\varphi + \theta_{y0}(t)) \end{bmatrix} => A_r(t) \begin{bmatrix} \cos(\theta_r(t)) \\ \sin(\theta_r(t)) \end{bmatrix},$$

which is achieved by the decoupling controller parameters

$$g_w = \frac{A_r(t)}{A_{y0}(t)}, \quad (8)$$
$$\varphi = \theta_r(t) - \theta_{y0}(t). \quad (9)$$

The designed decoupling controller is a time varying system and it can be implemented on the LLRF system. A simple time-invariant decoupling controller can be obtained if the signals are sampled and held at a flat-top instant $t_{FT}$. In this case, the decoupling controller is given by

$$W(s) = g_w(t_{FT}) \begin{bmatrix} \cos(\varphi(t_{FT})) & -\sin(\varphi(t_{FT})) \\ \sin(\varphi(t_{FT})) & \cos(\varphi(t_{FT})) \end{bmatrix} \quad (10)$$

$$\varphi(t_{FT}) = \theta_r(t_{FT}) - \theta_{y0}(t_{FT})$$
$$g_w(t_{FT}) = \frac{A_r(t_{FT})}{A_{y0}(t_{FT})}.$$

Both time varying and time-invariant decoupling controllers are implemented on the LLRF system and its mode is switched by the EDM's buttons (box 2 in figure 5).

## 4. PI FEEDBACK CONTROLLER AND CLOSE LOOP CALIBRATION

In the LANSCE LLRF system, independent, external instruments (AD5511 RF Envelope Detector, AD8302 RF Phase Detector) are deployed to measure the cavity field amplitude and the cavity field phase from cavity RF signal with respect to the 201.25MHz, 805MHz reference RF. The reference values of them (target values) are obtained during the beam tuning. These independent amplitude and phase values are used valuably when the operating points of the LLRF, high power RF subsystems, resonance control/water cooling systems, and other systems of LINAC are changed. The calls of the MATLAB functions (**E02_Calib.m, E03_Calib.m**) for the gain and phase calibrations are implemented on EPICS EDM screen with Shell Command buttons (box 1 in figure 5).

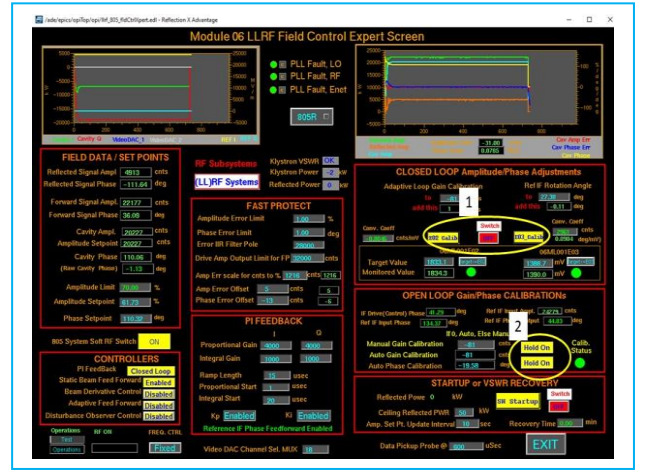

Figure 5. Field Control EDM Screen where Shell Command Buttons call **E02_Calib.m** and **E03_Calib.m** (block 1).

## 5. BEAM FEEDFORWARD COTNROLLER TUNING

Two Tuning Algorithms are implemented with MATLAB functions (**sbffgrid.m** for the grid search tuning and **sbffgradsearch.m** for the iterative gradient search tuning). **sbffgrid.m** can be run standalone iteratively with finer grids or it can be used as a parameter initialization process of the **sbffgradsearch.m**. A batch m-script(**bffctune_bat.m**) implements the menu that multiplexes function call to **sbffgrid.m, sbffgradsearch.m**. The call to the MATLAB functions is implemented on EPICS EDM screen (Figure 6) with Shell Command.

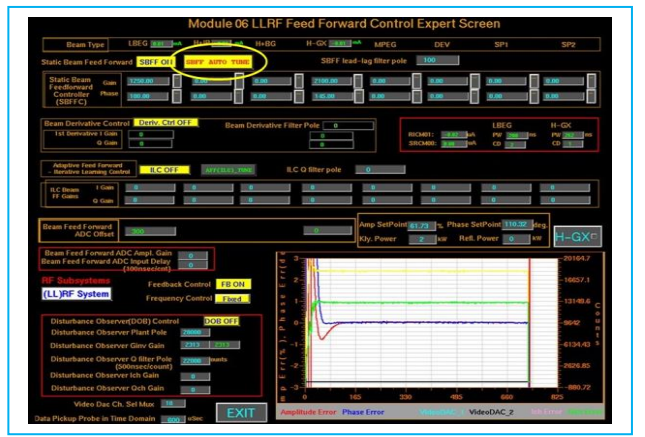

Figure 6. LLRF Feedforward Control EDM Screen where Shell Command Button calls **bffctune_bat.m.**

### Grid Search Based Feedoforward Controller Tuning

At a given beam current, the optimal controller gain and phase are searched on the discrete grids on the gain and phase to minimize the performance index (Table 1).

The gain grid is usually coarse and linked to the magnitude of the beam current. For each discrete monotonically increasing magnitude of beam current, the grid point of gain is determined so that the amplitude and the phase errors are well below the error requirements, $\pm 1.0\%$ amplitude error and $\pm 1.0$ degrees phase error. Experience shows that a single over-valued gain works well for both a small magnitude beam current and a full strength production beam current (8.0 mA). Figure 7 shows the phase grid search results where the single gain of 2500 counts were used for 4 different magnitudes of beam current. At the magnitude of beam current 3.2 mA, the lower end $\theta_L$ and the upper end $\theta_H$ of the phase grids are -180 degrees and +180 degrees, respectively, and the grid interval $\Delta\theta$ is 30 degrees. As the trial number is increased, the grid becomes finer, in which the grid interval update law is given in the Table 1.

Table 1. Summary of Phase Grid Search Algorithm

| Phase grid | $\overrightarrow{\rho_\theta} = [\theta_L:\Delta\theta:\theta_H]$ |
|---|---|
| Signals used | $e_A(t)$, $e_\theta(t)$ : errors in Amp/Phase Domain |
| Controller parameters | $\rho = [\rho_A \quad \rho_\theta]^T$: Amp/Phase Domain<br>$\rho_I = \rho_A \cos(\rho_\theta)$,<br>$\rho_Q = \rho_A \sin(\rho_\theta)$: I/Q Domain |
| Performance Index | $\overrightarrow{J_p} = [J_g(0), J_g(1), \ldots, J_g(M)]$,<br>$J_g(i) = max([\|e_A\|_\infty, \|e_\theta\|_\infty])$<br>i: $grid\ index$ |
| Phase grid update law | $\theta_L(j+1) = \rho_\theta^*(j) - \Delta\theta(j)$,<br>$\theta_H(j+1) = \rho_\theta^*(j) + \Delta\theta(j)$,<br>$\Delta\theta(j+1) = \frac{\Delta\theta(j)}{3}$,<br>j: $trial\ number$ |

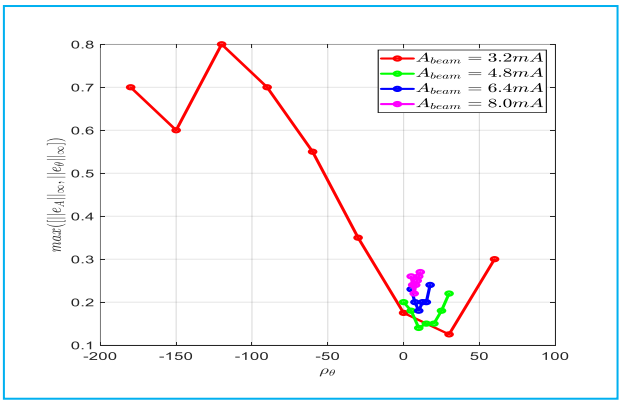

Figure 7. The feedforward controller tuning results of 4 trials of grid search. The experiment was performed at the LLRF low power teststand. As the beam current is increased, the phase grid becomes finer and for each beam current, the optimal controller phase found, $\rho_\theta^* = [30,10,10,7]$degrees. $\rho_A^* = 2500$ for all trials.

### Fine Tuning of the Beam Feedforward Controller using Disturbance Observer (DOB) estimates of complex beam current

When the magnitude of the beam current is full strength and the amplitude error and the phase error are within the error requirements, which is achieved by applying the grid search algorithm with the coarse phase grid interval, a fine tuning of the feedforward controller parameters are performed by applying the gradient search algorithm for the finite number of the collected data, iteratively. The objective of the fine feedforward controller tuning is to find the optimal value $\rho^*$ of the parameter $\rho = [\rho_I \quad \rho_Q]^T$ that minimizes the performance index (cost function) defined $J(\rho)$ defined as,

$$J(\rho) = \frac{1}{2N}\sum_{i=1}^{N}\left(e_I^2(i) + e_Q^2(i)\right) \quad (11)$$

where $e_I(i)$ and $e_Q(i)$ are the complex error components. This feedforward controller gain tuning problem is solved iteratively using a gradient search algorithm [2,4]

$$\rho^{(k+1)} = \rho^{(k)} - \alpha\,R^{-1}\,\nabla J[\rho^{(k)}] \quad (12)$$

where $\alpha$ is a scalar parameter to control the step size, $\rho^{(k)}$ is the parameter estimate in the $k^{th}$ iteration, $R$ is a matrix to modify the search direction, and $\nabla J[\rho^{(k)}]$ is the gradient of the cost function $J(\rho)$ with respect to the controller parameters $\rho^{(k)}$ of the $k^{th}$ iteration, which is obtained by

$$\nabla J[\rho^{(k)}] = \frac{\partial J}{\partial \rho}[\rho^{(k)}] = \begin{Bmatrix}\frac{\partial J}{\partial \rho_I}[\rho^{(k)}]\\ \frac{\partial J}{\partial \rho_Q}[\rho^{(k)}]\end{Bmatrix} \quad (13)$$

$$\frac{\partial J}{\partial \rho_I} = \frac{-1}{N}\sum_{i=1}^{N} e_I(i)\frac{\partial y_I(i)}{\partial u_I}\hat{d}_I(i) \quad (14)$$

$$\frac{\partial J}{\partial \rho_Q} = \frac{-1}{N}\sum_{i=1}^{N} e_Q(i)\frac{\partial y_Q(i)}{\partial u_Q(i)}\hat{d}_Q(i). \quad (15)$$

In (14) and (15), the complex beam current estimates $\hat{d}_I(i)$ and $\hat{d}_Q(i)$ obtained from DOB are used for the calculation of gradients. In addition, the derivatives $\frac{\partial y_I(i)}{\partial u_I}$ and $\frac{\partial y_Q(i)}{\partial u_Q(i)}$ are required to be computed. However, in our LLRF system, as a result of the decoupling controller $W(s)$, both are unity. Table 2 summarizes the data and the relevant equations coded in the MATLAB function **sbffgradsearch.m**.

## SUMMARY

In this paper, the procedure of the RF commissioning and beam operation of the LANSCE LINAC is addressed. Each step of the procedure can performed either manually or automatically. When the automatic process is performed, the developed MATLAB functions support the process. The automatic process is specially useful to manipulate multiple LANSCE LLRF system simultaneously.

Table 2. Summary of Gradient Search Algorithm

| Signals used | $e_I(t)$, $e_Q(t)$ : errors in I/Q domain |
|---|---|
| | $\hat{d}_I(t)$, $\hat{d}_Q(t)$ : complex beam currents estimate by DOB |
| Controller parameters | $\rho = [\rho_I \quad \rho_Q]^T$: Amp/Phase Domain |
| | $\rho_I = \rho_A \cos(\rho_\theta)$, $\rho_Q = \rho_A \sin(\rho_\theta)$: I/Q Domain |
| Performance Index | $J(\rho) = \frac{1}{2N}\sum_{i=1}^{N}(e_I^2(i) + e_Q^2(i))$<br>N: $sample\ number$ |
| Parameter Update Law | $\rho^{(k+1)} = \rho^{(k)} - \alpha \nabla J(\rho^{(k)})$,<br>k: $iteration\ number$ |
| Jacobian | $\nabla J(\rho^{(k)}) = \begin{bmatrix} -\frac{1}{N}\sum_{i=1}^{N} e_I(i)\,\hat{d}_I(i), \\ -\frac{1}{N}\sum_{i=1}^{N} e_Q(i)\,\hat{d}_Q(i) \end{bmatrix}$ |

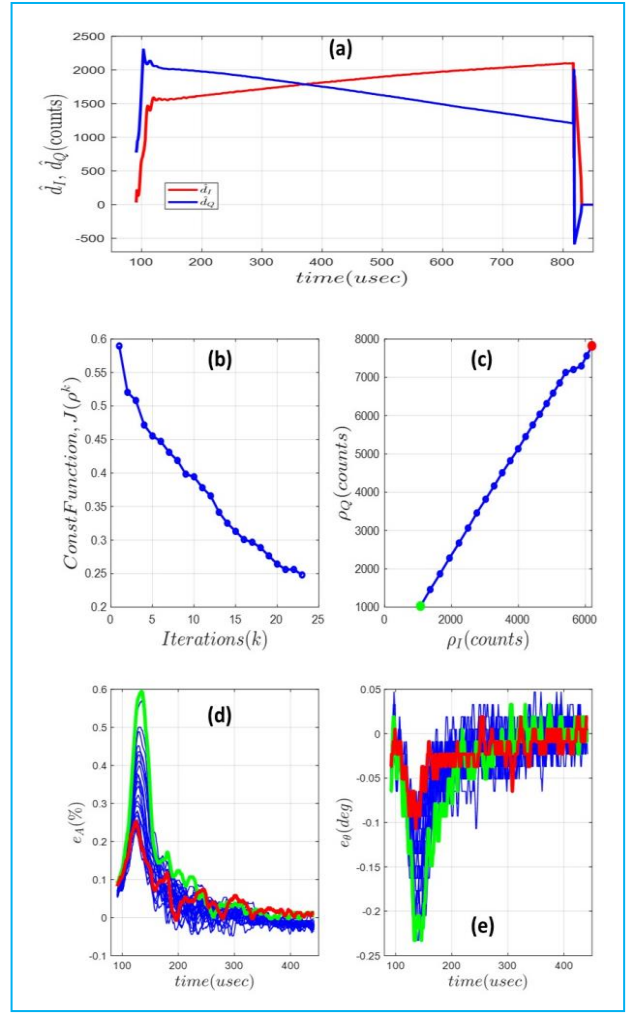

Figure 8. Performance results of the iterative tuning method: (a) Complex beam current estimates obtained from DOB; (b) Cost function; (c) Controller parameters; (d) Amplitude error; and (e) Phase error.